\documentclass{ws-ijmpa}

\begin{document}

\markboth{Savely G. Karshenboim}
{Precision study of positronium: testing bound state QED}

\catchline{}{}{}

\title{PRECISION STUDY OF POSITRONIUM: \\
TESTING BOUND STATE QED THEORY}

\author{\footnotesize SAVELY G. KARSHENBOIM\footnote{email: sek@mpq.mpg.de}}

\address{D. I. Mendeleev Institute for Metrology, 198005 St. Petersburg,
  Russia\\ 
Max-Planck-Institut f\"ur Quantenoptik, 85748 Garching, Germany}

\maketitle

%\pub{Received (Day Month Year)}{Revised (Day Month Year)}

\begin{abstract}
As an unstable light pure leptonic system, positronium
is a very specific probe atom to test bound state QED. 
In contrast to ordinary QED for free
leptons, the bound state QED theory is not so well understood and 
bound state approaches deserve highly accurate tests. We present a
brief overview of precision studies of positronium paying special 
attention to uncertainties of theory as well as comparison of theory
and experiment. We also consider in detail advantages and
disadvantages of positronium tests compared to other QED experiments.

\end{abstract}

\section{Introduction}

Quantum electrodynamics (QED) is the only quantum field theory which
can be successfully applied to a broad range of effects 
(bound states, scattering, decay) and energies from 
microwave radiation to high energies in the GeV range and deliver us 
various accurate predictions for measurable quantities with an
uncertainty reaching the ppt level.

QED of photons and leptons is an absolutely correct theory in a sense 
that its Lagrangian is well defined and in principle there is no
problem for performing any calculations.
\begin{itemlist}
 \item However, that is not sufficient since exact calculations are strongly
limited by increasing difficulties in the calculation of 
higher-order effects and we always have to deal with a finite
number of terms in the perturbative expansion. A question
therefore arises how to estimate terms, which are too complicated 
to be calculated and sometimes that is in part art.
 \item However, QED is in some way incomplete since electromagnetic 
interactions may involve hadrons and strong interactions which cannot 
be calculated {\em ab initio}. The weak interactions may also be 
involved, but usually it is not a problem to find related contributions.
 \item However, there is a basic theoretical problem while performing 
comparison to experiment. Theory is not in position to predict any
numbers. What theory can only do is to express some measurable
quantity in terms of others. In particular, to produce any
quantitative prediction within bound state QED, we need to be first able 
to determine with a proper accuracy values of basic fundamental
constants (the Rydberg constant $R_\infty$, the fine structure
constant $\alpha$ as well as different masses and magnetic moments)
and auxiliary parameters (such as the proton charge radius) due to 
the hadronic sector, which are necessary input data for bound state 
QED calculations.  
\item However, there is no way to solve the bound state problem in
general. The application of QED to the bound state forms a field
called {\em bound state QED}, which experiences its own difficulties, 
additionally to QED problems for free particles. 
\end{itemlist}

\section{Bound state QED}

Bound state QED is quite attractive as a training field to solve 
the bound state problem in quantum theory. It may be helpful for 
few-nucleon nuclei and for hadronic particles. In particular, 
there is a similarity in physics of positronium and quark-antiquark 
systems (mesons).

The difficulties of precision QED calculations for free particles 
are mainly due to an increasing number (up to one thousand) of
complicated diagrams (the four-loop level). The bound state QED theory 
deals with much simpler diagrams, however, the charged particles are
bound there rather than free, and thus the Coulomb exchange may be not 
a small effect. A detailed review on QED calculations for light atoms can be 
found in \cite{report}.

The free QED involves only one small parameter $\alpha$, while the
bound state QED theory needs at least three and all three expansions
are not quite good\cite{icap}.
\begin{itemlist}
 \item Indeed, we still have to deal with $\alpha$, the power of which 
indicates the number of QED loops involved. The expansion is
asymptotical but that is not important since the bound state
calculations mainly need one-loop and two-loop contributions, with 
three-loop effects being important rather seldom.
 \item The Coulomb strength $Z\alpha$ appears because of binding
effects and the parameters $\alpha$ and $Z\alpha$ behave in a quite 
different way. There is a number of contributions where we need to sum
over an infinite number of Coulomb exchanges (e.g. the Bethe
logarithm). Importance of all the exchanges assumes their essentially 
non-relativistic behaviour and still allows a $Z\alpha$ expansion, 
which cannot be avoided since  calculations exact in $Z\alpha$  are 
possible for a few contributions only. However, that is not a well
behaving expansion, because the limit $Z\alpha\to0$ is related to an 
unbound two-body system and thus leads to a non-analytic behaviour of 
perturbative expressions. 
The non-analyticity in Coulomb systems is usually accompanied with numerous
logarithmic factors. 
For $Z=1$ (hydrogen, muonium, positronium) one can find: 
$\ln(1/Z\alpha) \simeq 5$. The corrections known up to now may include
up to cube of this logarithm, $\ln^3(1/Z\alpha) \simeq 120$ 
(for the Lamb shift), and up to logarithm squared for hyperfine
structure and positronium physics\cite{log1}. 
For several reasons  the non-logarithmic 
contributions also involve big coefficients\cite{icap}. 
 \item A bound state problem supposes that we deal with an atom 
consisting of an orbiting particle(s) and an attractor (nucleus) and 
that involves one more parameter, a ratio $m/M$ of masses of the 
orbiting particle (mainly an electron) and the nucleus which for 
conventional atoms is $m/M\simeq 0.5\cdot 10^{-3} A^{-1}$, 
for muonium is $m/M\simeq 1/207$ and for positronium is $m/M=1$. 
The behaviour of the expansion in $m/M$ is not good since the limit
$m/M\to 0$ is related to a kind of bound ``neutrino". The non-analytic 
behaviour shows itself in logarithmic terms and, e.g., 
for muonium $\ln(M/m)\simeq 5.5$. 
 \item Some more parameters are involved due to nuclear effects, such
as, e.g., a ratio of the Bohr radius to the nuclear radius.
\end{itemlist}

Thus, the bound state QED theory involves a rather rich spectrum of 
problems and it deserves to be tested, particularly in spite of the
lack of well established universal prescriptions appropriate 
for the two-body bound state problem in general. There are two basic
problems of the precision bound state QED theory: Lamb shift and 
hyperfine structure.
\begin{itemlist}
 \item The Lamb shift calculations mainly need an external field 
approximation, while recoil corrections are less important and only
the simplest of them are involved.
 \item In contrast, calculations of the hyperfine interval are crucially 
affected by the recoil effects, while some external field effects 
(e.g., the higher-order two-loop corrections) are relatively less 
important. 
\end{itemlist}

\section{Hyperfine structure in bound state QED}

Most of interest to positronium properties is due to recoil effects 
which are crucial since $m/M=1$. For this reason we consider here in 
more detail studies of the hyperfine structure in light atoms.

\begin{figure}[htbp] 
\centerline{ \psfig{file=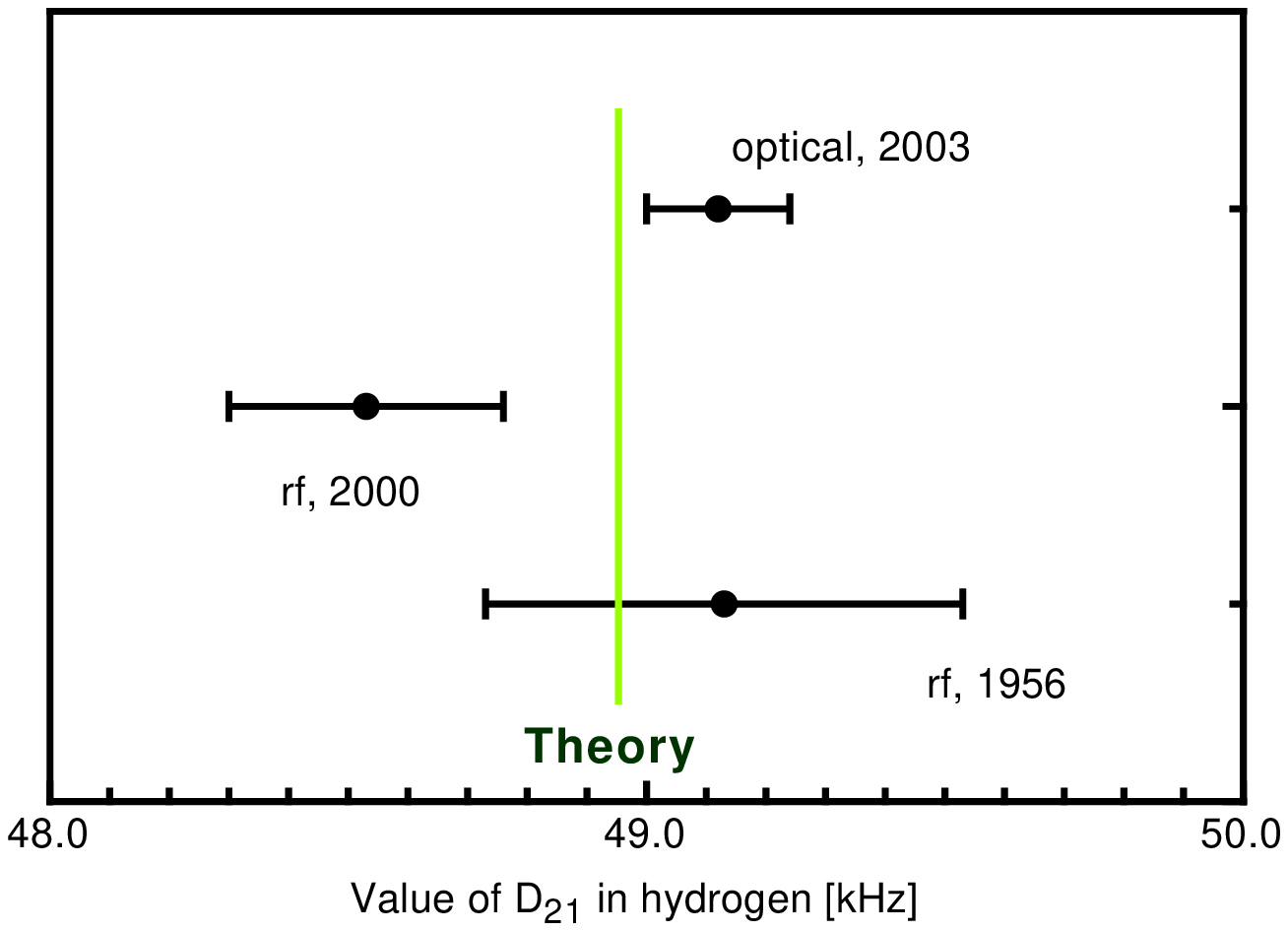,width=6cm}~~~~~~~~~~~~\psfig{file=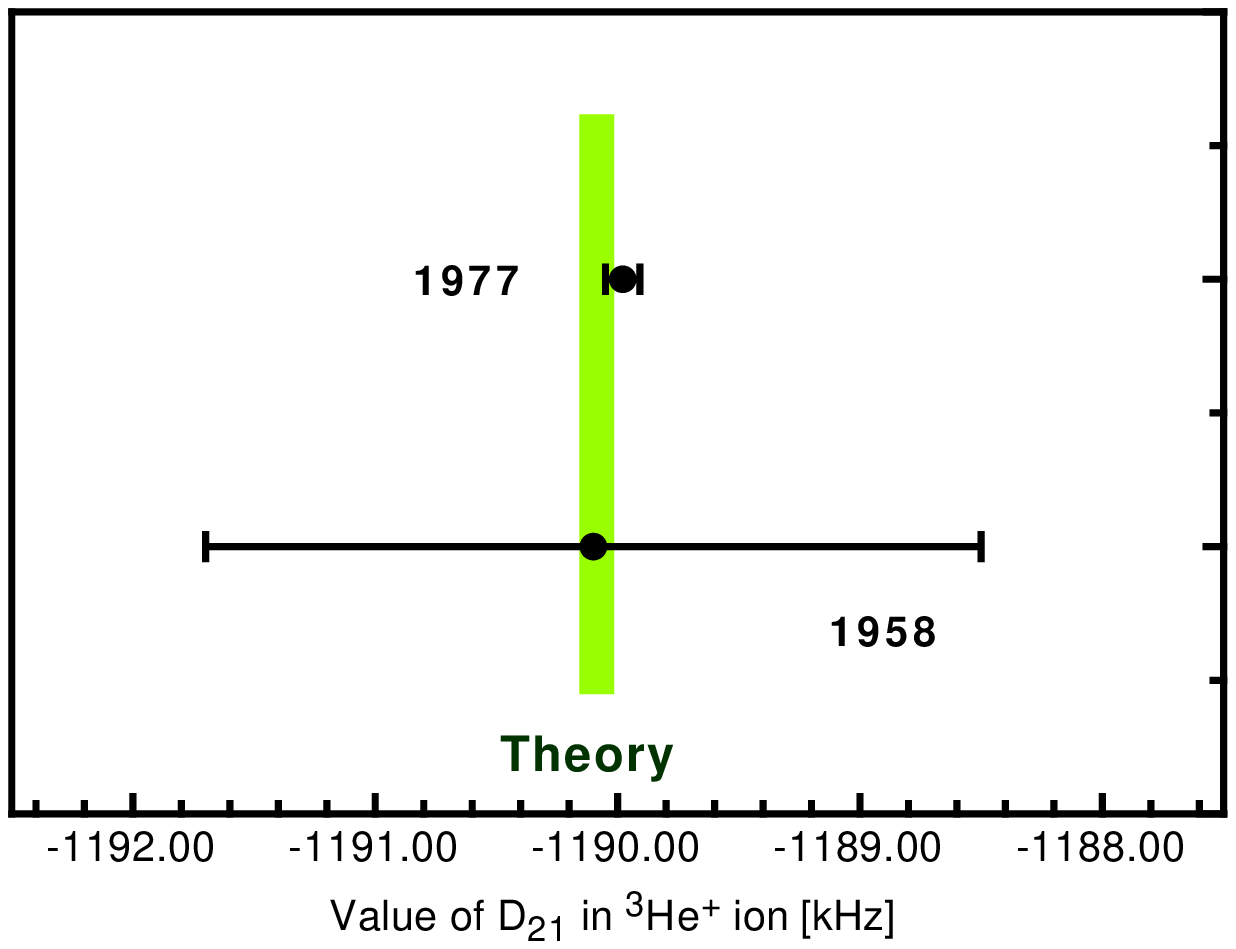,width=6cm}}
\vspace*{8pt}
\caption{Determination of the $D_{21}$ difference in hydrogen and 
helium-3 ion. The references can be found in Table 10.\label{f:d21}}
\end{figure}

Magnetic effects are relativistic effects and thus, in contrast to the 
Coulomb interaction responsible for the Lamb shift, the higher
momentum transfers and shorter distances are more important. At  
shorter distances the recoil and nuclear-structure effects are
enhanced. The nuclear effects in hydrogen and other light atoms
dominate over the bound state QED\cite{new_rep,d21}. Still there are 
three possible QED tests with the hyperfine structure in which the  
problem of nuclear effects can be avoided.
\begin{itemlist}
 \item A comparison of the $1s$ and $2s$ hyperfine intervals that 
offers a specific difference\cite{new_rep,d21}
\begin{equation}
D_{21} = 8 \nu_{\rm HFS}(2s) - \nu_{\rm HFS}(1s) \;,
\label{e:d21}
\end{equation}
which is immune to leading effects of the nuclear structure.
 \item A comparison of conventional and muonic atoms for the same nucleus.
 \item A study of a pure leptonic atomic system such as muonium and 
positronium.
\end{itemlist}
Determination of the $2s$ hyperfine interval in muonic hydrogen is in part the
goal of an PSI experiment\cite{psi} which is now in progress.  
The HFS interval in the $1s$ and $2s$ states was successfully studied
for several light atoms. The more complicated measurement is related
to the metastable $2s$
state\cite{exph2s,rothery,2shydr,expd2s,exphe2s,prior}. $D_{21}$
theory is compared to experiment in Fig.~\ref{f:d21}. The theory 
of $D_{21}$ for several atoms is summarized in Table \ref{t:d21}. 
The dominant uncertainty of QED theory is due to higher-order one-loop and 
two-loop corrections in order $\alpha(Z\alpha)^3$ and
$\alpha^2(Z\alpha)^2$ and recoil corrections in order 
$\alpha(Z\alpha)^2(m/M)$ (in units of the $1s$ HFS splitting). 
The higher-order nuclear effects also substantially contribute to 
the uncertainty.

\begin{table}[hbtp]
\tbl{Theory of the specific difference $D_{21}$ in light hydrogen-like atoms$^{5,4}$. 
%\protect\cite{d21,new_rep}
\label{t:d21}}
{\begin{tabular}{@{}lccc@{}} %\toprule
\hline
~Contribution & Hydrogen & Deuterium & $^3$He$^+$ ion\\
& [kHz] & [kHz] & [kHz] \\
\hline
~$D_{21}({\rm QED3})$ & 48.937 &  11.305\,6 & -1\,189.252\\
~$D_{21}({\rm QED4})$ & {0.018(3)} & {0.004\,3(5)}  &-1.137(53)\\
~$D_{21}({\rm nucl})$ & {-0.002} & {0.002\,6(2)} & 0.317(36)\\
\hline
~$D_{21}({\rm theo})$ & 48.953(3) &  11.312\,5(5) & -1\,190.083(63)  \\
\hline
\end{tabular}}
\end{table}

\begin{table}[hbtp]
\tbl{
Theory of the $1s$ hyperfine splitting in muonium. The calculations$^{13}$
%\cite{hamu1} 
have been adjusted to $\alpha^{-1}= 137.035\,998\,76(52)$$^{14}$ 
%\cite{kinoalpha} 
and $\mu_\mu/\mu_p=3.183\,345\,17(36)$ which was obtained from the
analysis of the data on Breit-Rabi levels in muonium$^{15}$
%\cite{MuExp} 
and other less accurate experiments. 
\label{t:mu}}
{\begin{tabular}{@{}lc@{}} %%\toprule
\hline
~~~Term  &  ~~~~~~~~~~~~~~$\Delta E$~~~~~~~~~~~~~~ \\
&   [kHz] \\
\hline
~~~$E_F$ & 4\,459\,031.88(50)~~ \\ 
~~~$a_e$ & 5\,170.926(1)~\\
~~~QED2  &- 873.147~~~~\\
~~~QED3  &- 26.410~~\\
~~~QED4  &~~~~~~- 0.551(218)\\
~~~Hadronic~~~ &~~~~~0.240(4)\\
~~~Weak  &- 0.065~~\\
\hline 
~~~Total & ~~~4\,463\,302.73(55)\\
\hline
\end{tabular}}
\end{table}

The muonium theory  of the $1s$ hyperfine interval is summarized in 
Table~\ref{t:mu}.
The dominant uncertainty of QED theory\cite{hamu1} is due to 
higher-order recoil corrections in order $\alpha(Z\alpha)^2(m/M)$ and 
$(Z\alpha)^3(m/M)$ which are also in part responsible for the
uncertainty of $D_{21}$ (see above). The other important part of the
uncertainty is related to the determination of the leading term (so-called
Fermi energy) 
\begin{equation}
\frac{E_F}{h} = \frac{16}{3} \,\alpha^2 \,\frac{\mu_\mu}{\mu_B}\,R_\infty c
\left(\frac{m_\mu}{m_\mu+m_e} \right)^3 
\end{equation}
in terms of fundamental constants $\mu_\mu/\mu_B$ and $\alpha$ and inaccuracy
in their determination. 

\section{Why positronium ?}

As mentioned before, the recoil effects are better seen in
positronium. Below we consider the positronium spectrum and a
comparison of theory to experiment. We find that the uncertainty in 
calculating all experimentally studied  transitions is related to the 
same recoil contributions as for the muonium HFS interval. 

The recoil effects play a crucial role in a two-body bound problem
showing how closely the bound system is to a real two-body system. 
However, the significance of positronium is not limited by the possibility 
to verify the theory of recoil corrections.
\begin{itemlist}
 \item Since the corrections of interest are enhanced ($m/M$ is not a 
suppressing factor any longer), the fractional accuracy for successful 
high-precision tests is now relatively low. As a result, in contrast
to hydrogen, the interpretation of the measurements of the $1S-2S$
interval does not crucially involve knowledge of the Rydberg constant 
with high accuracy. A study of the hyperfine interval does not require 
a value of the fine structure constant with high accuracy as it is in 
muonium. Since $m/M=1$, it is not necessary to determine either $m/M$ 
or $\mu_{\rm Nucl}/\mu_B$ in an additional experiment. In other
words, positronium offers several high precision tests of bound state
QED without determinations of fundamental constants with
high accuracy.
 \item The hyperfine and recoil effects are enhanced and thus can be
seen not only in a direct study of the hyperfine structure but also
in the investigation of the gross or fine structure in contrast to 
hydrogen and muonium. Thus, positronium offers a few transitions which 
can be studied with high accuracy (see Table~\ref{t:pos}). Adding to 
that an opportunity of different experiments on positronium
annihilation, we find a big variety of properties to be studied.
\item It is not even necessary to mention that as a light pure 
leptonic atomic system, the positronium atom is free of hadronic
effects. It is important that positronium is light because 
hadronic effects in leptonic systems involve a high momentum
transfer. They can be seen in muonium and in the muon anomalous
magnetic moment they are responsible for a dominant part of the 
theoretical uncertainty, while they are strongly suppressed for 
positronium and the anomalous magnetic moment of electron.
\end{itemlist}

Since positronium is a very specific atom, the notation for
positronium is slightly different from other two-body atoms. First, 
in two-body atoms, even in hydrogen and muonium, it is customary to
keep the value of the nuclear charge $Z$ in order to recognize exchange 
photons and photons of QED radiative effects and thus to trace the 
origin of different corrections. In positronium there is an
interference between both kinds of photons because of annihilation 
diagrams and thus it is meaningless to keep $Z$. 

Hydrogen and other two-body atoms are one-electron systems and one can 
use for them both small (e.g., $1s$) and capital (e.g., $1S$) 
letters to denote levels. The former are used for an electron, while
the latter are for all electrons in an atom and that is indeed the
same for single-electron atomic systems. Here, we prefer to use
small-case letters for hydrogen and others. Since the nuclear spin
effects are not suppressed, positronium has a structure of energy
levels (in respect to their spin and angular momentum) rather similar 
to a two-electron system (such as the neutral helium atom) and one has 
to use only capital letter for its orbital momentum.

\begin{table}[hbtp]
\tbl{\label{t:pos} Theoretical predictions for positronium. Comparison 
to experimental data is presented in the figures quoted in the last column 
of the table.}
{\begin{tabular}{@{}lcc@{}} %\toprule
\hline
~~Quantity & Prediction & Figure \\
\hline 
~~$\Delta\nu_{\rm HFS}(1S)$ & 203\,391.7(6) MHz& \protect\ref{f:HFS} \\
$\Delta\nu(1^3S_1-2^3S_1)$  & 1\,233\,607\,222.2(6) MHz& \protect\ref{f:1s2s}\\
\hline
~~$\Delta\nu(2^3S_1-2^3P_0)$  & 18\,498.25(9) MHz & \protect\ref{f:2s2p} \\
~~$\Delta\nu(2^3S_1-2^3P_1)$  & 13\,012.41(9) MHz & \protect\ref{f:2s2p}\\
~~$\Delta\nu(2^3S_1-2^3P_2)$  & 8\, 625.70(9) MHz & \protect\ref{f:2s2p}\\
~~$\Delta\nu(2^3S_1-2^1P_1)$  & 11\,185.37(9) MHz & \protect\ref{f:2s2p}\\
\hline 
~~$\Gamma ({\rm p\!-\!Ps})$ & 7989.62(4)  $\mu$s$^{-1}$ & \protect\ref{f:pps}\\
~~$\Gamma ({\rm o\!-\!Ps})$ & 7.039\,96(2)  $\mu$s$^{-1}$ &
\protect\ref{f:ops} \\ 
~~${\rm Br}_{4\gamma/2\gamma}({\rm pPs})$ & $1.439(2)\cdot10^{-6}$ &
\protect\ref{f:pps}\\ 
\hline
\end{tabular}}
\end{table}

We summarize theoretical predictions for various transitions and decay 
rates in Table~3. The Table does not contain experimental results but
only references to figures where a comparison of theory to experiment 
is performed. Our results for decay rates and the HFS interval are 
slightly different from those quoted in the literature because we include into
the decay rates contributions related to the decay to non-minimal number of 
photons (four for parapositronium and five for orthopositronium) which 
are sometimes omitted and because of our conservative estimation of
uncertainty. To be conservative, we estimate the final 
uncertainty by half of the value of the leading logarithmic terms, if the 
non-leading term is unknown or reduces the logarithmic contribution, 
and by half of the value of the whole logarithmic contribution, if the
non-leading term enhances the leading term (cf. [16]).

\section{Positronium spectrum}

%\subsection{The $1S$ hyperfine interval}

\begin{figure}[htbp] 
\centerline{\psfig{file=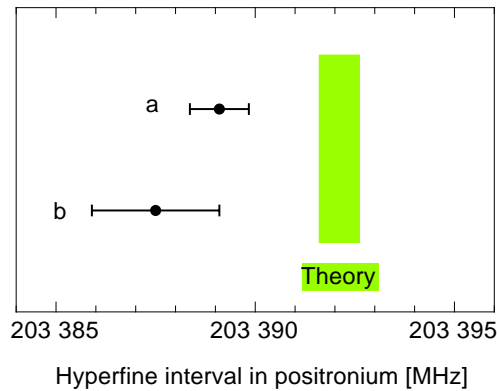,width=7cm}}
\vspace*{8pt}
\caption{The positronium $1S$ hyperfine splitting: a comparison of theory to
  experiment ($a$ -- [17], $b$ -- [18]).\label{f:HFS}} 
\end{figure}

Let us now consider the positronium spectrum in more detail.
The hyperfine interval in positronium has been measured with the
highest absolute accuracy among positronium transitions. A comparison
of theory and experiment is presented in Fig.~\ref{f:HFS} while theory
is summarized in Table~\ref{t:HFS}. The theoretical contributions are 
classified in units of the Fermi energy 
\begin{equation}
E_F =\frac{7}{12}\,\alpha^4mc^2
\;,
\end{equation}
which is defined including the virtual one-photon annihilation. E.g.,
QED2 is related to corrections of relative order $\alpha^2$. This
notation is helpful for a comparison to the theory of decay rates, while
the absolute units, also presented in the Table, allow a simple and 
direct comparison with theory of the $1S-2S$ interval. Agreement with 
experiment is not perfect (within 2.5-3 standard deviations) which 
demonstrates that positronium is not very well understood.

\begin{table}[hbtp]
\tbl{QED contributions to the $1S$ hyperfine interval in positronium. The
  uncertainty is presented following [16]. 
%\cite{pos}.
\label{t:HFS}}
{\begin{tabular}{@{}lr@{}} %\toprule
\hline
~~Term~~~~~~&$\Delta E$~~~~~~~~~~~~~~~~~~~~ \\
&[MHz]~~~~~~~~~~~~~~~~~~ \\
\hline
~~$E_F$~~~~~~~$\alpha^4mc^2$ & 204\,386.6~~~~~~~~~~~~~~~~\\
~~QED1~~~$\alpha^5mc^2$       & -1\,005.5~~~~~~~~~~~~~~~~\\
~~QED2~~~$\alpha^6mc^2$       & 11.8\protect\cite{posqed2}~~~~~~~~~~~~~\\
~~QED3~~~$\alpha^7mc^2$       &-
1.2(6)\protect\cite{log1,logps1,logps2,logps3}\\ 
\hline 
~~Total & ~~~~203\,391.7(6)~~~~~~~~~~~~\\
\hline
\end{tabular}}
\end{table}

%\subsection{The $1S-2S$ two-photon transition} 

\begin{figure}[htbp] 
\centerline{\psfig{file=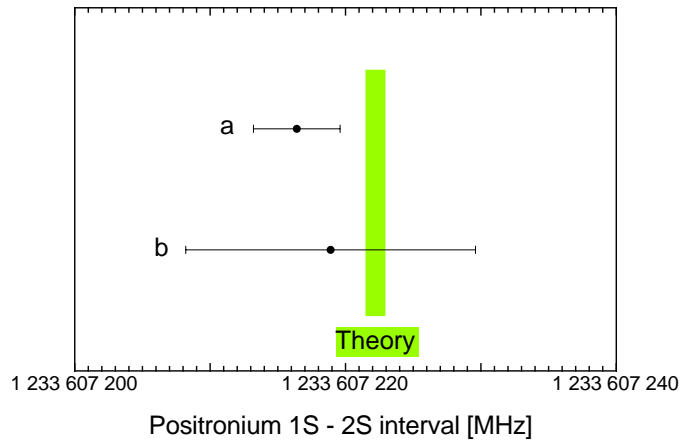,width=9cm}}
\vspace*{8pt}
\caption{Theory and experiment ($a$ -- [25], $b$ -- [26]) for determination of
  the $1^3S_1-2^3S_1$ interval in positronium.\label{f:1s2s}} 
\end{figure}

The $1S-2S$ interval was measured in orthopositronium by a method of 
three-photon ionization which has a resonance related to the two-photon 
excitation of the metastable $2S$ state from the ground state. The
absolute uncertainty of the $1S-2S$ interval is compatible with 
the hyperfine splitting although somewhat lower. The uncertainty of the 
theoretical predictions for both hyperfine interval and the 
$1S-2S$ transition (see Table~\ref{t:1s2s}) is determined by 
a correction of order $\alpha^7mc^2$. That is related to $\alpha^3
E_F$ in the positronium hyperfine interval ($QED3$) and thus to 
$QED4$ corrections in atoms with heavy nucleus ($m/M\ll1$), namely, 
$\alpha(Z\alpha)^2(m/M)E_F$ and $(Z\alpha)^3(m/M)E_F$. A comparison of 
theory to experiment is presented in Fig.~\ref{f:1s2s}.

\begin{table}[hbtp]
\tbl{Theory of the $1^3S_1-2^3S_1$ interval in positronium.\label{t:1s2s}}
{\begin{tabular}{@{}lr@{}} %\toprule
\hline
~~Term~~~~~~&$\Delta E$~~~~~~~~~~~~ \\
&[MHz]~~~~~~~~~~ \\
\hline
~~$\alpha^2mc^2$ &  1\,233\,690\,735.1~~~~~~~~\\
~~$\alpha^4mc^2$ &  -82\,005.6~~~~~~~~\\
~~$\alpha^5mc^2$ &  -1\,501.4~~~~~~~~\\
~~$\alpha^6mc^2$ &  -7.1\protect\cite{pk1}~~~~~\\
~~$\alpha^7mc^2$ &  1.2(6)\protect\cite{pk2} \\
\hline 
~~Total & 1\,233\,607\,222.2(6)~~~~\\
\hline
\end{tabular}}
\end{table}

%\subsection{The $n=2$ fine structure}

\begin{figure}[htbp] 
\centerline{\psfig{file=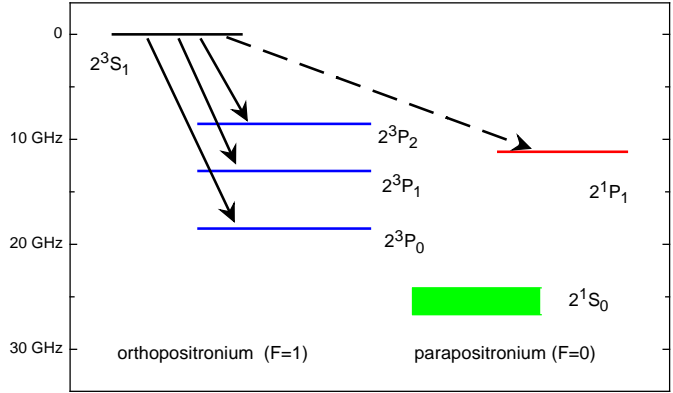,width=9cm}}
\vspace*{8pt}
\caption{Fine structure in positronium. The scale does not allow to
see a width of all levels, but only of the broad $2S$ singlet state. 
The $2^3S_1-2^1P_1$ transition is forbidden, however, it may be
observed by applying a magnetic field.\label{f:fs}}
\end{figure}

\begin{figure}[htbp] 
\centerline{\psfig{file=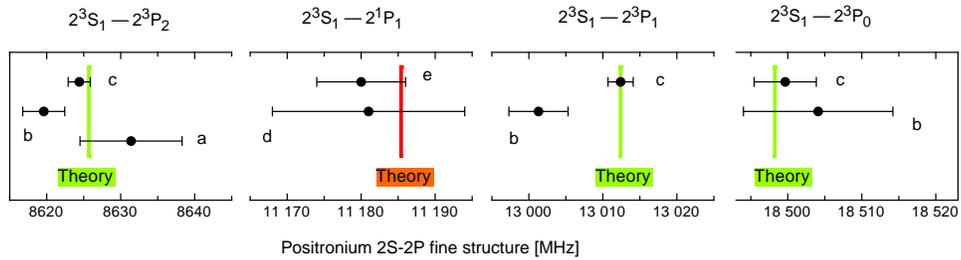,width=13cm}}
\vspace*{8pt}
\caption{Measurements of the fine structure transitions at $n=2$. Experiments
  were performed at Brandeis University ($a$ -- [27]), University of Michigan
  ($b$ -- [28] and $d$ -- [30]) and Mainz University ($c$ -- [29] and $e$ --
  [31]).\label{f:2s2p}} 
\end{figure}

The theoretical uncertainty for the $n=2$ fine structure in
positronium (see Fig.~\ref{f:fs}) is compatible with that for 
the hyperfine interval and the $1S-2S$ transition, but the experiment 
on $2S-2P$ transitions cannot provide competitive results 
(see Fig.~\ref{f:2s2p}). However, progress is possible\cite{conti}. 

\section{Positronium annihilation} 

Positronium is an unstable atom because of  electron and positron
annihilation. The $1S$ state of positronium decays mainly into two or 
three photons depending on its spin. Thus, the orthopositronium
($S=1$) annihilates into an odd number of photons, while the 
parapositronium ($S=0$) into an even number. The leading rates are
\begin{eqnarray}
\Gamma_{3\gamma}^{(0)}({\rm oPs}) = \frac{2(\pi^2-9)}{9\pi}\, \alpha^6
\frac{mc^2}{\hbar} \;, \\ 
\Gamma_{2\gamma}^{(0)}({\rm pPs}) = \frac{1}{2}\,\alpha^5 \frac{mc^2}{\hbar}
\;. 
\end{eqnarray}

%\subsection{Lifetime of orthopositronium in the ground state} 

\begin{figure}[htbp] 
\centerline{\psfig{file=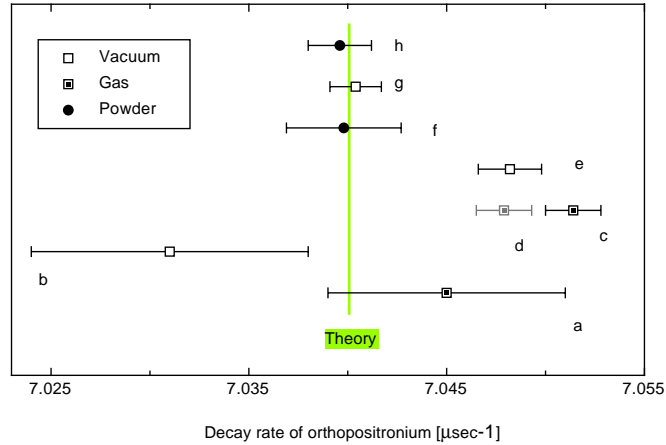,width=9cm}}
\vspace*{8pt}
\caption{Determination of the orthopositronium annihilation decay rate: a
  comparison of theory to experiment ($a$ -- [34],  
$b$ -- [35],
$c$ -- [32],
$d$ -- [36],
$e$ -- [33],
$f$ -- [37],
$g$ -- [38],
$h$ -- [39]).
\label{f:ops}}
\end{figure}

The lifetime of orthopositronium was responsible for a long-standing 
discrepancy of theory and most accurate experiments performed at 
the University of Michigan\cite{6c,6e} while theory was in fair
agreement with less accurate experiments, which disagreed with the 
Michigan data. The present situation is summarized in
Fig.~\ref{f:ops}. 
Hopefully, the crisis seems to be resolved since the new Michigan 
vacuum result\cite{6g} now agrees with theory and the 
reevaluation\cite{conti} of the former gas experiment\cite{6c} 
shifted the value towards theory. A further reexamination of systematic
effects is in progress 
and the final uncertainty will probably be higher\cite{6i}.

\begin{figure}[htbp] 
\centerline{\psfig{file=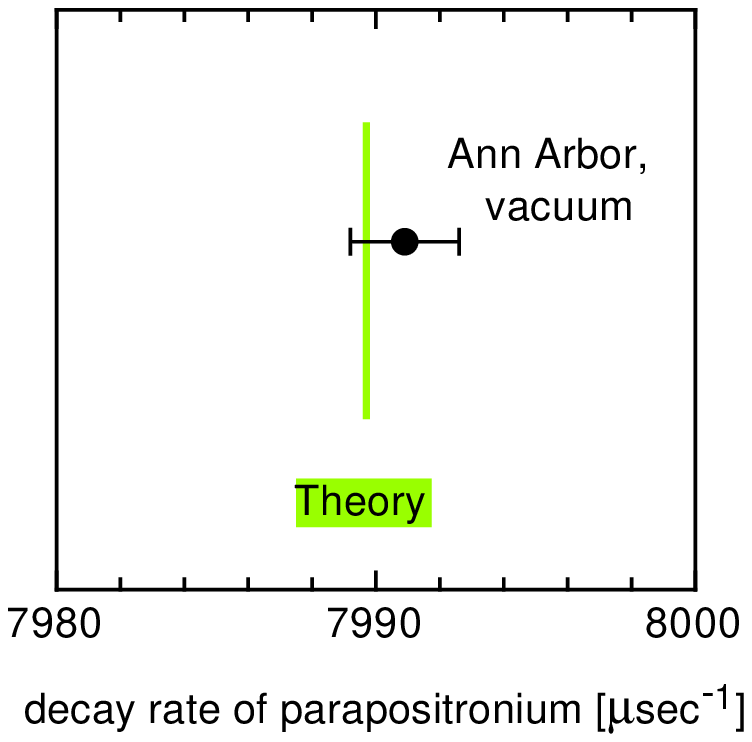,width=5cm}~~~~~~~~~~~~~~~\psfig{file=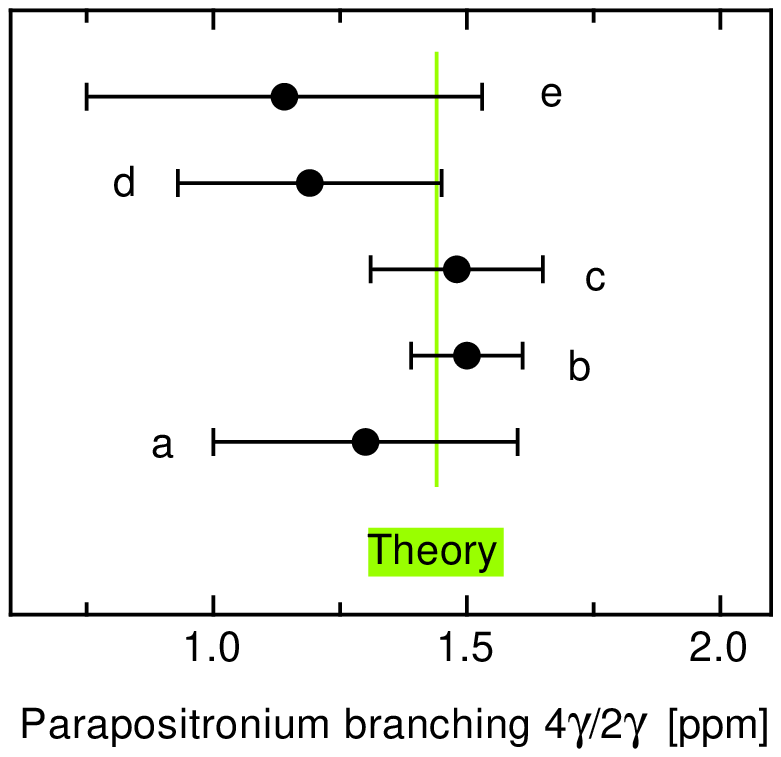,width=5cm}}
\vspace*{8pt}
\caption{Annihilation of parapositronium: decay rate measured in University of
  Michigan in Ann-Arbor (left) and branching Br($4\gamma/2\gamma$) measured
  around the world ($a$ -- [53], $b$ -- [54], $c$ -- [55], $d$ -- [56], $e$ --
  [48]).\label{f:pps}} 
\end{figure}

Meanwhile, the lifetime of parapositronium also measured at the 
University of Michigan\cite{7} is in fair agreement with the 
theoretical prediction (see Fig.~\ref{f:pps}). The theoretical
predictions for the annihilation decay rates for ortho- and parapositronium
are summarized in Table~\ref{t:gam}. 

%\subsection{Lifetime of parapositronium in the ground state}

\begin{table}[hbtp]
\tbl{Theory of annihilation decay rate of ortho- and parapositroium (the $1S$
  state). The leading contributions are defined above in Eqs. (4) and (5). The
  decay rate of ortho/parapositronium into five/four photons is included into
  QED2 terms. 
\label{t:gam}}
{\begin{tabular}{@{}lll@{}} %\toprule
\hline
~~ Contribution & ~~~~Decay rate & ~~~~Decay rate \\
&of orthopositronium &of parapositronium \\
&~~~~~~[$\mu$sec$^{-1}$] & ~~~~~~[$\mu$sec$^{-1}$]  \\
\hline
~~$\Gamma^{(0)}$                 &~7.211\,17              &8\,032.50\\
~~$\alpha\cdot\Gamma^{(0)}$      &-0.172\,30              &~ \,-47.25 \\
~~$\alpha^2\cdot\Gamma^{(0)}$    &~0.001\,11(1)\cite{afs}  &~~\,~4.43(1)\cite{cmy1}\\
~~$\alpha^3\cdot\Gamma^{(0)}$    &-0.000\,01(2)\cite{log1,declog} & ~~\,-0.08(4)\cite{log1,declog} \\
\hline 
~~Total &  7.039\,96(2) & 7989.62(4) \\
\hline
\end{tabular}}
\end{table}

We do not consider here exotic decay modes related to possible new
physics, but rather rare modes possible within QED and the Standard
Model. Two rare QED modes are allowed at a detectable level. Their theory and
experiments are  
summarized in Table~\ref{t:rar}. We also include in the Table a 
photonless annihilation of orthopositronium into a pair 
of neutrino-antineutrino. The other modes with a pair of 
$\nu\overline{\nu}$ are also possible but their branching ratios are a few 
orders of magnitude lower.

\begin{table}[hbtp]
\tbl{Rare positronium decay modes. The theoretical uncertainty for $4\gamma$
  decay is related to unknown higher-order corrections.\label{t:rar}} 
{\begin{tabular}{@{}cccc@{}} %\toprule
\hline
~Rare mode & Leading mode & Branching (theor)  & Branching (exp)  \\
\hline
~${\rm oPs} \to 5\gamma$ 
 & ${\rm oPs} \to 3\gamma$ 
 & $1.0\cdot10^{-6}$, Ref. \protect\cite{gamt,5gamt} 
 & $\left(2.2^{+2.6}_{-1.8}\right)\cdot10^{-6}$, Ref. \protect\cite{5game}~~\\
&& & $1.7(11)\cdot10^{-6}$, Ref. \protect\cite{45gam}~~\\
\hline
~${\rm oPs} \to \nu\overline{\nu}$ & ${\rm oPs} \to 3\gamma$
&$6.2\cdot10^{-21}$, Ref. \protect\cite{nunu} & $<5.8\cdot10^{-4}$ (90\%
CL)\protect\cite{noth1}~~\\ 
&&& $<2.8\cdot10^{-4}$ (90\% CL)\protect\cite{noth2}~~\\
\hline
~${\rm pPs} \to 4\gamma$ & ${\rm pPs} \to 2\gamma$ & $1.439(2)\cdot10^{-6}$
Ref. \protect\cite{4gamt}  
   & $1.30(30)\cdot10^{-6}$, Ref. \protect\cite{4game1}~~\\
&& & $1.50(11)\cdot10^{-6}$, Ref. \protect\cite{4game2}~~\\
&& & $1.48(17)\cdot10^{-6}$, Ref. \protect\cite{4game3}~~\\
&& & $1.19(26)\cdot10^{-6}$, Ref. \protect\cite{4game4}~~\\
&& & $1.14(39)\cdot10^{-6}$, Ref. \protect\cite{45gam}~~\\
\hline
\end{tabular}}
\end{table}

\begin{figure}[htbp] 
\centerline{ \psfig{file=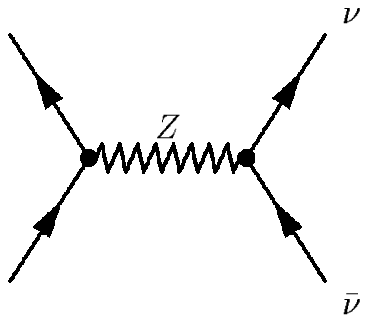,width=3cm}~~~~~~~~~~~~\psfig{file=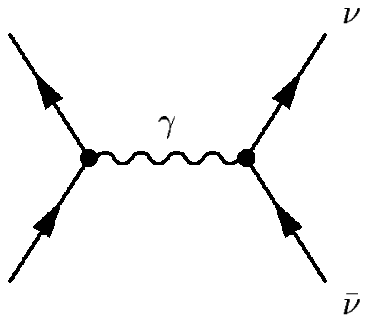,width=3cm}}
\vspace*{8pt}
\caption{Annihilation of orthopositronium into a pair of
$\nu\overline\nu$: 
via $Z$-boson (left) and photon (right). The former diagram is related 
to the Standard Model, while the latter is present 
only if $\mu_\nu\neq 0$.\label{f:onu}} 
\end{figure}

We have to note that the branching fraction of the orthopositronium
decay into neutrinos is so low, that this decay is unlikely to be soon 
detected,
however, a mode $oPs \to {\it nothing}$ (since the neutrino is not
detectable) can still be of interest if the neutrino has a non-vanishing 
magnetic moment (see, e.g., Fig.~\ref{f:onu}). This problem was discussed 
in part in \cite{weak} but we do not like to present here any
numbers. In our opinion, results of such analysis can depend on a model
introducing neutrino mass and magnetic moment. Still, studies of the 
pure neutrino modes can provide a limit on the $\tau$-neutrino 
magnetic moment at a level above (but possibly not much above) the current 
limits\cite{lims}. Since systematic effects are different and 
the interpretation depends on the model, we think it may be important
to have several independent limitations.

\section{Positronium and other QED tests}

Tests with positronium are quite different from other tests. 
Advantages and disagvantages of positronium studies and some other 
QED experiments are listed in Table~\ref{t:adv}. 

\begin{table}[hbtp]
\tbl{Advantages and disadvantages of different QED tests for conventional
  light atoms, pure leptonic bound systems and free leptons.\label{t:adv}} 
{\begin{tabular}{@{}lll@{}} %\toprule
\hline
~System & Value & Dominant uncertainties \\
\hline
~Hydrogen & Lamb shift & Nuclear size, higher-order two-loop effects,
$R_\infty$\\ 
~Hydrogen & $1s$ HFS & Huge uncertainty due to nuclear structure \\
~Hydrogen & $D_{21}$ & Experiment\\
~$^3{\rm He}^+$ & $D_{21}$ & Higher-order two-loop, recoil, nuclear effects,
experiment\\ 
~$^4{\rm He}^+$ & Lamb shift & Nuclear size, higher-order two-loop effects,
experiment\\ 
\hline
~Muonium & $1s$ HFS & Higher-order recoil effects, $\mu_\mu/\mu_B$, $\alpha$\\
~Positronium & $1S$ HFS & Higher-order recoil effects, experiment\\
~Positronium & $1S-2S$ & Higher-order recoil effects, experiment\\
~Positronium & $2S-2P$ & Experiment\\
~Orthopositronium & $\Gamma(1s)$ & Experiment\\
~Parapositronium & $\Gamma(1s)$ & Experiment\\
\hline
~Electron & $g-2$ & Uncertainty: $\alpha$, cavity QED effects\\
~Muon & $g-2$ & Uncertainty: hadronic effects\\
\hline
\end{tabular}}
\end{table}

To compare different QED tests with/without  positronium, we also
check which contributions are crucial for a comparison of theory to 
experiments. They are summarized in Table~\ref{t:ord}. 

\begin{table}[hbtp]
\tbl{Comparison of the bound state QED theory and experiment: crucial orders
  of  magnitude for the energy levels and decay rates needed for the
  comparison$^2$ (in units of $mc^2$).\label{t:ord}} 
{\begin{tabular}{@{}lc@{}} %\toprule
\hline
Atom, value & Crucial order(s) \\
 & [$mc^2$] \\
\hline 
Hydrogen (gross structure) & $\alpha(Z\alpha)^7$, $\alpha^2(Z\alpha)^6$ \\
Hydrogen (fine structure)              & $\alpha(Z\alpha)^7$, $\alpha^2(Z\alpha)^6$ \\
Hydrogen (Lamb shift)              & $\alpha(Z\alpha)^7$, $\alpha^2(Z\alpha)^6$ \\
$^3$He$^+$ ion ($2s$ HFS)  & $\alpha(Z\alpha)^7m/M$, $\alpha^2(Z\alpha)^6m/M$, $\alpha(Z\alpha)^6(m/M)^2$, $(Z\alpha)^7(m/M)^2$\\
$^4$He$^+$ ion (Lamb shift)              & $\alpha(Z\alpha)^7$, $\alpha^2(Z\alpha)^6$ \\
Muonium ($1s$ HFS)      & $\alpha(Z\alpha)^7m/M$, $\alpha(Z\alpha)^6(m/M)^2$, $(Z\alpha)^7(m/M)^2$ \\
\hline
Positronium ($1S$ HFS)  & $\alpha^7$ \\
Positronium ($1S-2S$)      & $\alpha^7$ \\
Positronium ($2S-2P$)       & $\alpha^7$ \\
Para-positronium (decay rate)       & $\alpha^7$ \\
Ortho-positronium (decay rate)      & $\alpha^8$ \\
Para-positronium ($4\gamma$ branching)  & $\alpha^8$ \\
Ortho-positronium ($5\gamma$ branching) & $\alpha^8$ \\
\hline
\end{tabular}}
\end{table}

\section{Higher-order logarithmic corrections and uncertainty of
positronium calculations}

To consider HFS tests in detail, we summarize the results on a study
of the hyperfine structure in light two-body atoms in
Table~\ref{t:HFSQED}. In these tests the QED uncertainty has never
been a limiting factor for comparison of theory and experiment.
The following uncertainties may also be involved: due to nuclear
effects, determination of fundamental constants or measurement. One of
them is always bigger than the theoretical uncertainty. 

The theory and experiment are in general in good agreement. The
uncertainty for the $1s$ hyperfine interval in muonium and positronium 
and in part of $D_{21}$ for helium-3 ion is related to the same recoil 
corrections in order $\alpha^3(m/M)E_F$ (see Table~\ref{t:ord}). 
The positronium uncertainty in fractional units is much higher than
that for experiments with heavy atoms ($m/M\ll1$), however, its
sensitivity to higher-order recoil effects is at approximately the same level.

The higher-order recoil contributions are known only in the
logarithmic approximation and we estimate the uncertainty as half
the value of the leading term if the next-to-leading term is unknown 
or cancels a part of the leading term. However, if 
the leading and the next-to-the leading terms are of the same sign,
 we estimate the uncertainty as a half value of the whole logarithmic 
contribution (cf. \cite{log2,hamu1,pos}). Thus, a calculation of the 
next-to-leading terms does not reduce an uncertainty. A reason for such 
a conservative estimation is that the leading logarithmic term is 
mostly a result of a single 
contribution without any cancellation. For the $ns$ state it is 
state-independent. For these reasons it has a kind of natural value
and can be used to estimate the non-leading terms. In contrast, 
the non-leading term has quite accidental value and may be sometimes 
quite below the natural level. Thus, it cannot be used alone to
estimate properly the 
next term of the logarithmic expansion. However, we cannot simply 
ignore the next-to-leading term, particularly in the case when it is 
not too small. In the present paper we follow a compromise which 
allows to achieve a conservative estimation of the uncertainty. 
A calculation of the next of the non-leading terms helps us to check
the reliability of this estimation using only the leading term and meantime is
a necessary step towards a calculation of the  
whole contribution beyond the logarithmic approximation which is strongly
needed for muonium and positronium HFS. 

\begin{table}[hbtp]
 \tbl{Comparison of experiment and theory for hyperfine structure in 
 hydrogen-like atoms. In the  $D_{21}$ case the reference is given only for 
 the $2s$ hyperfine interval.\label{t:HFSQED}}
{\begin{tabular}{@{}lcccc@{}} %\toprule
\hline
~Atom  & ~~~~~~Exp.~~~~~~ & ~~~~~~Theory~~~~~ &  ~~~$\Delta/\sigma$~~~ &
~~~$\sigma/E_F$~~~ \\ 
 & [kHz] & [kHz]  &   & [ppm] \\
 \hline 
~Hydrogen, $D_{21}$  &  49.13(13)\protect\cite{2shydr} & 48.953(3)  & 1.4 &
0.09 \\  
~Hydrogen,  $D_{21}$  &  48.53(23)\protect\cite{rothery}  &  & -1.8 & 0.16 \\
~Hydrogen,  $D_{21}$  &  49.13(40)\protect\cite{exph2s} & & 0.4 &  0.28 \\
~Deuterium, $D_{21}$  &  11.16(16)\protect\cite{expd2s}  & 11.312\,5(5) & -1.0
& 0.49 \\ 
~$^3$He$^+$ ion, $D_{21}$~~~  &-1\,189.979(71)\protect\cite{prior}
&-1\,190.083(63) &1.10 &0.01  \\ 
~$^3$He$^+$, $D_{21}$ & -1\,190.1(16)\protect\cite{exphe2s} &  &  0.0 & 0.18 \\
\hline 
~Muonium, $1s$ & 4\,463\,302.78(5)\protect\cite{MuExp}& 4\,463\,302.88(55)&
-0.18 &0.11\\ 
~Positronium, $1S$ & 203\,389\,100(740)\protect\cite{1} & 203\,391\,700(600) &
-2.9 &4.4\\ 
~Positronium, $1S$ & 203\,397\,500(1600)\protect\cite{2}& & -2.5 &8.2\\
\hline
\end{tabular}}
\end{table}

\section{Soft and hard QED effects}

Not only essential orders of QED corrections offer a possibility of 
comparing the efficiency of different experiments. The bound state QED 
theory clearly recognizes two kinds of contributions: the soft-photon 
contribution and the hard-photon contribution. The latter are very
similar to free QED, while the former essentially involve binding
effects. There are two most important soft-photon contributions.
\begin{itemlist}
\item Crucial corrections in the external field approximation are due
to the higher-order two-loop self-energy. The inaccuracy in its
calculation determines an uncertainty of the Lamb shift calculations 
for hydrogen and hydrogen-like ions, while similar effects involving
the magnetic field significantly contribute to the uncertainty of 
$D_{21}$ in the helium-3 ion. 
\item Recoil corrections are crucially important for tests with hyperfine
structure and they determine the uncertainty of muonium HFS, 
positronium spectrum (hyperfine interval, $1s-2s$ transition, fine
structure) and a part of the uncertainty of $D_{21}$ for the helium-3 ion. 
\end{itemlist}

Effects of the hard-photon exchange are very similar to effects of
free QED. We note, however, that the regions of integration in 
momentum space are different. The most accurate free QED calculations 
which may be compared to experiment are related to the anomalous
magnetic moments of electron and muon. The integration for them is 
performed over a kind of isotropic region in Euclidean space 
($\vert k_0\vert\sim\vert{\bf k}\vert$) and the crucial level is the 
four-loop approximation. For the bound state problems there are two other
specific regions of integration.
\begin{itemlist}
\item For some problems it is sufficient to apply an external field 
approximation to hard-photon corrections 
and thus loops of exchange photons are related to zero energy transfer 
($k_0=0$). Even for recoil effects the 
integration is essentially not covariant in Euclidean space 
including a contribution from a specific region $\vert k_0\vert\ll\vert{\bf
  k}\vert$.  
The highest crucial orders are related to 
four-loop corrections for the external-field approximation and
three-loop corrections for recoil effects. In contrast to the
anomalous magnetic moment, that is a calculation for two different 
particles with 
its own simplifications and difficulties.  
\item The other specific situation for integration is related to the 
positronium annihilation when some photons are real 
($k_0=\vert{\bf k}\vert$). The orthopositronium studies 
(main decay mode and branching fraction for five-photon decay) allow to 
check calculations of four-loop corrections in such a non-isotropic
region of integration. 

The accuracy of the branching fraction 
determination for the four-photon decay of parapositronium is at the
level of a few percent. Because of a relatively large $\alpha$-contribution to
the $4\gamma$ annihilation rate\cite{4gamt} 
\begin{equation}
\Gamma({\rm pPs}\to4\gamma) = \Gamma^{(0)}({\rm pPs}\to4\gamma)\times
\left(1-14.5(6)\frac{\alpha}{\pi}\right)
\end{equation}
the experiments are approaching a level, where four-loop diagrams are
important. 
\end{itemlist}

Thus, we see that the bound state problems supply us with a
possibility to check modern algorithms for four-loop calculations and 
that is competitive to the anomalous magnetic moment of electron.
We note, however, that the difficulties which appear at the four-loop 
level are a large number of digrams with high-dimension integrations 
and overlapping UV divergencies. Neither annihilation nor exchange 
loops are ultraviolet divergent, but the exchange loops involve 
strong infrared divergency.

One can also compare\cite{yf} soft-QED effects of the virtual
one-photon annihilation and the real two- and three-photon
annihilation. Soft Coulomb corrections to the hard annihilation block 
can be presented in fractional units. A crucial theoretical
uncertainty related to $\alpha^3$ corrections (QED3 term) is at the level of
few ppm, which can be hardly achieved for a measurement of the annihilation
rates. 

\section{Summary} 

Thus, we summarize our paper with the following statements.
\begin{itemlist}
\item Positronium spectroscopy offers a reliable test of our 
understanding of higher-order recoil corrections within bound state QED, 
which play a significant role in other QED tests with the hyperfine
structure. The theoretical uncertainty for positronium is of a pure 
QED origin and other effects such as the nuclear structure effects or 
inaccuracy in determination of fundamental constants are not
involved. Studies of recoil effects are of particular interest because 
their significance answers a question whether we study a really
two-body system. In conventional atoms (hydrogen, helium) the role of
the nucleus as a particle is reduced. The positronium is a truly 
two-body system which is the closest to the neutral helium atom. 
In contrast to two-body atoms, the QED uncertainty in helium is bigger 
than that of experiment and theory needs a significant improvement.  
Thus, positronium theory serves as an intermediate step between 
hydrogen and helium. 
\item A study of positronium annihilation including rare decay modes 
allows to test approaches to the calculation of four-loop diagrams 
but for a specific integration region. Today the 
four-loop approximation is the highest level for precision
calculations of any measurable QED quantities (if indeed no big
logarithms are involved such as those in scattering kinematics).
\item There are exotic decay modes of positronium beyond QED, but within the
  Standard Model 
and in particular a decay of orthopositronium into a pair of neutrino and
antineutrino, which is a dominant mode beyond QED.  
The branching fraction is very low and it cannot be detected
presently. Still, a study of this channel can probably give a
limit for the magnetic moment of the $\tau$ neutrino which, although 
 somewhat weaker than the current limits, will, however,
have completely different systematic effects.
\end{itemlist}

\section*{Acknowledgements}

The author is grateful to A. Badertscher, R. Conti, A. Czarnecki, S. Eidelman,
M. Felcini,  
D. Gidley, S. Gninenko, R. Ley and A. Rubbia for useful and stimulating 
discussions. This work was supported in part by RFBR under grants 
\# 02-02-07027 and \# 03-02-16843.

\end{document}